\documentstyle[aps,epsfig]{revtex}

\begin{document}

\title{Study   of   Dissipative   Collisions   of   $^{20}$Ne   ($\sim$7-11
MeV/nucleon) + $^{27}$Al}

\author   {Aparajita  Dey,  C.~Bhattacharya,  S.~Bhattacharya,  T.~K.~Rana,
S.~Kundu, K.~Banerjee, S.~Mukhopadhyay, S.~R.~Banerjee, D.~Gupta, R.~Saha}

\address{  Variable Energy Cyclotron Centre, Sector - 1, Block - AF, Bidhan
Nagar, Kolkata - 700 064, India.}

\maketitle

\begin{abstract}
The  inclusive  energy distributions of complex fragments (3 $\leq$Z $\leq$
9) emitted in the reactions $^{20}$Ne (145, 158, 200, 218 MeV) +  $^{27}$Al
have   been  measured  in  the  angular  range  10$^{o}$  -  50$^{o}$.  The
fusion-fission and the deep-inelastic components of the fragment yield have
been extracted using multiple  Gaussian  functions  from  the  experimental
fragment  energy  spectra.  The  elemental  yields  of  the  fusion-fission
component have been found to be fairly well exlained in  the  framework  of
standard  statistical  model.  It is found that there  is strong competition 
between the fusion-fission and the deep-inelastic processes at these energies.  
The time scale of the deep-inelastic process
was estimated  to  be  typically  in  the  range  of  $\sim$  10$^{-21}$  -
10$^{-22}$  sec.,  and  it  was  found to decrease with increasing fragment
mass.  The  angular  momentum   dissipations   in   fully   energy   damped
deep-inelastic process have been estimated from the average energies of the
deep-inelastic components of the fragment energy spectra. It has been found
that, the estimated angular momentum dissipations, for lighter fragments in
particular, are more than those predicted by the empirical sticking limit.

\end{abstract}

PACS number(s): 25.70.Jj, 24.60.Dr, 25.70.Lm

\section{Introduction}
Complex  fragment  emission  in heavy-ion induced reactions involving light
nuclei ($A_{target} + A_{projectile} \lesssim 60$) at  bombarding  energies
well  above  the  Coulomb barrier has been studied quite extensively in the
recent                                                                years
\cite{sa99,sa87,c90,be1,be2,be3,r91,sh82,s87,s88,h94,f96,a93,b2,be4,c98} to
understand  the  origin  of  fragment  emission  and the role of underlying
dynamics. It is well known that  different  types  of  reaction  mechanisms
contribute  to fragment emission at different energy regions. At bombarding
energies near the Coulomb barriers, complete fusion  (CF)  process  is  the
dominant reaction mechanism. At higher energies, this process is limited by
the  contributions of other competing processes, such as quasi-elastic (QE)
and deep-inelastic (DI) collisions. The  CF  cross-section  increases  with
incident  energy  at  lower energies and reaches a near-saturation value at
higher  energies.  On  the  other   hand,   non-fusion   processes   become
increasingly  dominant  at  higher energies. Thus, the fragments emitted in
light heavy-ion collisions at energies well above the Coulomb  barrier  may
have different origins, which extend from partially relaxed processes, such
as, quasi-elastic (QE) collision / projectile break-up \cite{car89,pada90},
deep-inelastic          (DI)          transfer         and         orbiting
\cite{sh82,s87,bhat05,bh04,sh79,sh821}, to fully relaxed fusion-fission
(FF) \cite{moretto,sa91,Matsuse97,dha2,Szanto97,Szanto96} process. In  some
cases, the structure of the nuclei has also been found to play an important
role.  Therefore,  the  characterization  of  the origin of fragments is of
utmost importance to extract  information  on  the  relaxation  of  various
degrees  of  freedom (energy and angular momentum dissipation, for example)
in heavy ion collision in this energy domain. However, for  light  systems,
the  distinction  between different reaction mechanisms, FF and orbiting or
DI processes in particular, is very difficult as there is strong overlap in
the elemental distributions of the fragments emitted in these processes.

Binary  decay  of the light composite system $^{47}$V has been investigated
quite extensively in the past years. In some cases,  (where  the  composite
system  $^{47}$V  was produced through inverse kinematical reactions, like,
$^{35}$Cl  +  $^{12}$C  \cite{be1,be2,pirr97,be962},   $^{31}$P   +
$^{16}$O  \cite{r91},  $^{23}$Na  + $^{24}$Mg \cite{be3}) it was found
that the $^{47}$V composite system deexcites statistically. In these cases,
the emitted fragment  yields  show  1/$\sin\Theta_{c.m.}$  -  like  angular
dependence  and  have  angle-independent  mean  total  kinetic energy (TKE)
values in agreement with the decay of a fully energy equilibrated composite
system.  The  experimental  cross-sections  are  well  explained  with  the
predictions  of the extended Hauser-Feshbach method (EHFM) \cite{Matsuse97}
and thus suggest a fusion-fission origin. It  was  further  concluded  that
orbiting  process  \cite{s87}  does not play any significant role in the
decay of $^{47}$V composite system. On the other hand, studies on the  same
system,   produced   through  direct  kinematical  reactions  ($^{20}$Ne  +
$^{27}$Al \cite{kozub75,nato,van83}), showed that the angular distributions
of fully damped fragments are forward  peaked  and  fall  off  faster  than
1/$\sin\Theta_{c.m.}$,   which   are   characteristic   of   DI  processes.
Subsequently, assuming the fragment yield to be of DI origin (and  assuming
the  sticking  limit  for  the  angular  momentum  dissipation),  a  highly
elongated configuration for the $^{20}$Ne + $^{27}$Al di-nuclear system was
conjectured \cite{nato}).

It is clearly evident from the above that some degree of ambiguity prevails
over the interpretation of the fragment yield data in the decay of $^{47}$V
composite  system.  To resolve the ambiguity, it is necessary to understand
the roles played by various competing processes in this energy regime.  For
example,  there  is strong competition between FF and DI processes at these
energies,  which  should  be  decifered  properly  to  extract   meaningful
information  about  the  reaction  mechanism.  In  recent  years,  we  have
developed a scheme for the decomposition of FF and  DI  components  of  the
fragment  yield \cite{b2,bh04} in order to study systematically the
competition between FF and DI processes in light heavy  ion  collisions  at
energies  well  above the barrier; In this paper, we report an experimental
study of fragment emission in the decay of $^{47}$V composite system,
produced through $^{20}$Ne (145 -- 218 MeV) + $^{27}$Al reactions. Some part of the
$^{20}$Ne (145 MeV) + $^{27}$Al data has already been published \cite{bh04}.
The FF and DI components of the fragment yields have been extracted in each
case  to  study  the  systematics  of the two processes in the above energy
range.

The  paper  has  been  arranged as follows. The experimental procedures are
given in the next  section.  The  experimental  results  are  presented  in
Sec.~III. Finally, The discussion and conclusion are given in Sec.~IV.

\section{Experimental Procedures}

The  experiments  have  been  performed using the accelerated $^{20}$Ne ion
beams from the Variable Energy Cyclotron at Kolkata. The beam energies were
145, 158, 200 and 218 MeV; the target was made of self-supporting $^{27}$Al
of thickness $\sim$ 515 $\mu$g/cm$^2$. Fragments have been  detected  using
two  types  of  solid  state  telescopes;  telescopes with $\sim$ 10 $\mu$m
$\Delta$E [Si(SB)], 300 $\mu$m E  [Si(SB)]  were  used  to  detect  heavier
fragments  (5  $\leq$ Z $\leq$ 9), whereas telescopes with $\sim$ 10 $\mu$m
$\Delta$E [Si(SB)], 5 mm E [Si(Li)] were used for the detection of  lighter
fragments (3$\leq$ Z $\leq$ 5). The two types of telescopes were mounted on
two  arms of the scattering chamber which could move independently. Typical
solid angle subtended by each detector was $\sim$0.6  msr.  The  telescopes
were  calibrated  using  elastically  scattered  $^{20}$Ne  ion from Au, Al
targets and Th-$\alpha$ source. The systematic errors in the data,  arising
from the uncertainties in the measurements of solid angle, target thickness
and  the  calibration  of  current  digitizer  have  been  estimated  to be
$\approx$ 15\%.

\section{Experimental Results}

\subsection{Energy distribution}

\begin{figure}

{\epsfig{file=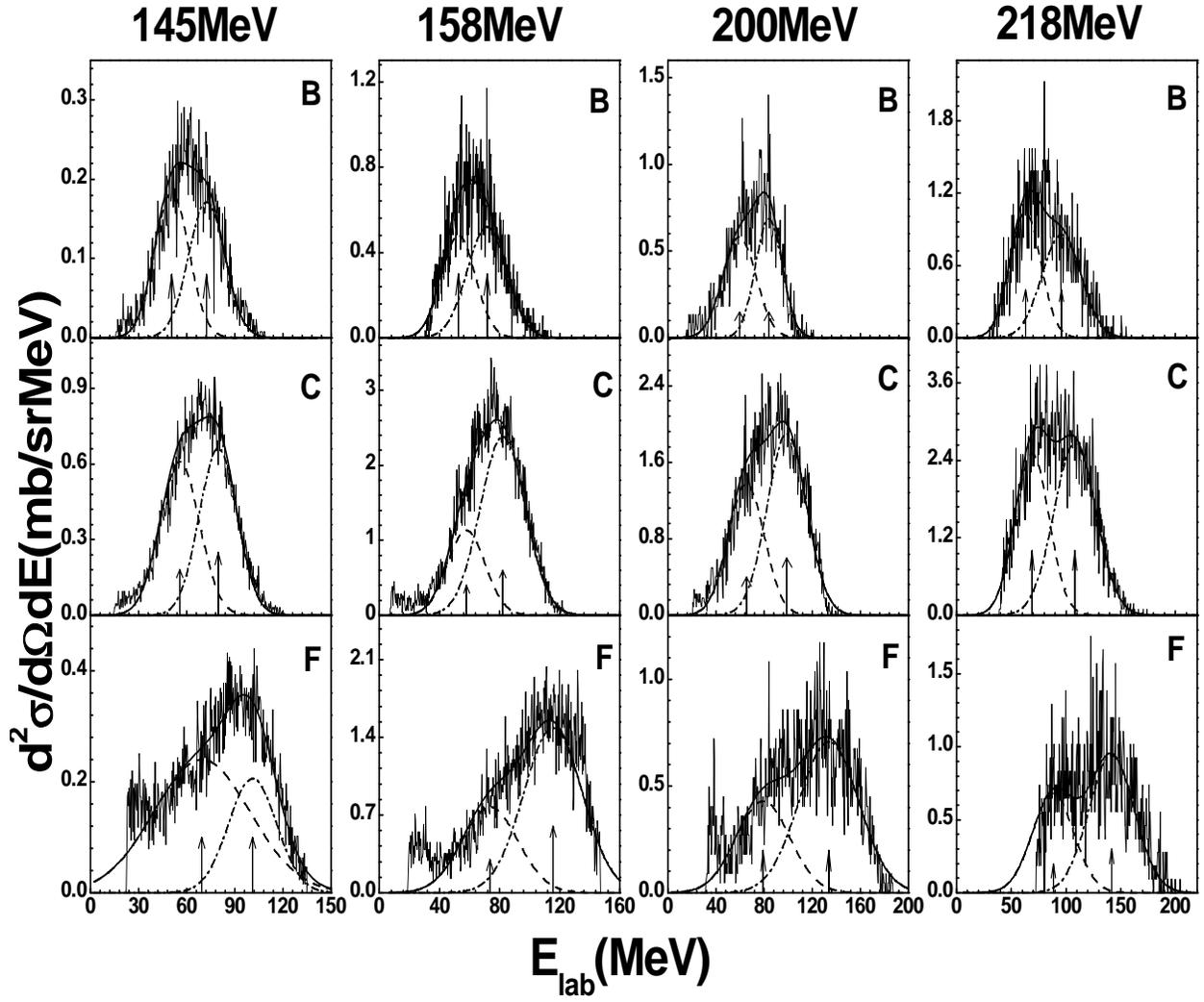,width=17.0cm,height=15.0cm}}
\caption{Inclusive energy distributions for different fragments emitted in $^{20}$Ne + $^{27}$Al reaction at
different bombarding energies at an angle $\theta_{lab}$ = 15$^{o}$. The FF, DI components and the sum
(FF + DI) are denoted by dashed, dash-dotted and solid curves, respectively. The arrows indicate the centroids
of the fitted Gaussians.}

\label{neal1}

\end{figure}

The  inclusive energy distributions have been measured for the fragments (3
$\leq$ Z $\leq$ 9) emitted in the reaction $^{20}$Ne  +  $^{27}$Al  at  the
bombarding energies 145, 158, 200 and 218 MeV,
respectively, in the angular range 10$^{o}$ -- 50$^{o}$. Typical
fragment  energy spectra (at $\theta_{lab}$ = 15$^{o}$) have been displayed
in Fig.~\ref{neal1} for different bombarding energies. It is  evident  from
Fig.~\ref{neal1}  that  the  shapes  of  the  energy spectra of the heavier
fragments  (viz.,  F)  are  quite  different  from  those  of  the  lighter
fragments,  viz.,  B  and C at all bombarding energies. It is mainly due to
the variation of the relative contributions of  DI  and  FF  processes  for
different  fragments.  The  contributions of FF and DI components have been
estimated by fitting the measured energy spectra with Gaussian functions as
per the procedure laid down in Ref.~\cite{b2}. The  energy  spectra  of
different  fragments  at  each  angle  have  been  fitted with two Gaussian
functions in two steps. In the first step, the FF component of the fragment
energy distribution has been extracted in the  following  way;  The  energy
distribution  of  the FF component was taken to be a Gaussian. The centroid
of the Gaussian was obtained from Viola  systematics  \cite{viola,be961},
adapted  for light nuclear systems \cite{be96}, of total kinetic energies
of mass-symmetric fission fragments duly corrected  for  asymmetric  factor
\cite{be4}.  The  FF  component of the energy spectrum thus obtained was
then subtracted from the full energy spectrum. In the  next  step,  the  DI
component  was  obtained  by  fitting the subtracted energy spectrum with a
second  Gaussian.  This  is  illustrated  in  Fig.~\ref{neal1},  where  the
extracted FF (dashed curve) and DI (dash-dotted curve) components have been
displayed  along with the experimental data for all bombarding energies. It
is clear from Fig.~\ref{neal1} that the experimental  energy  spectra,  for
all  fragments  at all bombarding energies, may be explained fairly well as
sum of two Gaussian functions representing the FF and DI components (sum of
FF and DI are shown by solid curve). In each spectrum, the arrow  at  lower
energy corresponds to the centroid of the Gaussian for the FF component and
the  arrow at higher energy corresponds to the centroid of the Gaussian for
the DI component.

\subsection{Angular distribution}
\begin{figure}

{\epsfig{file=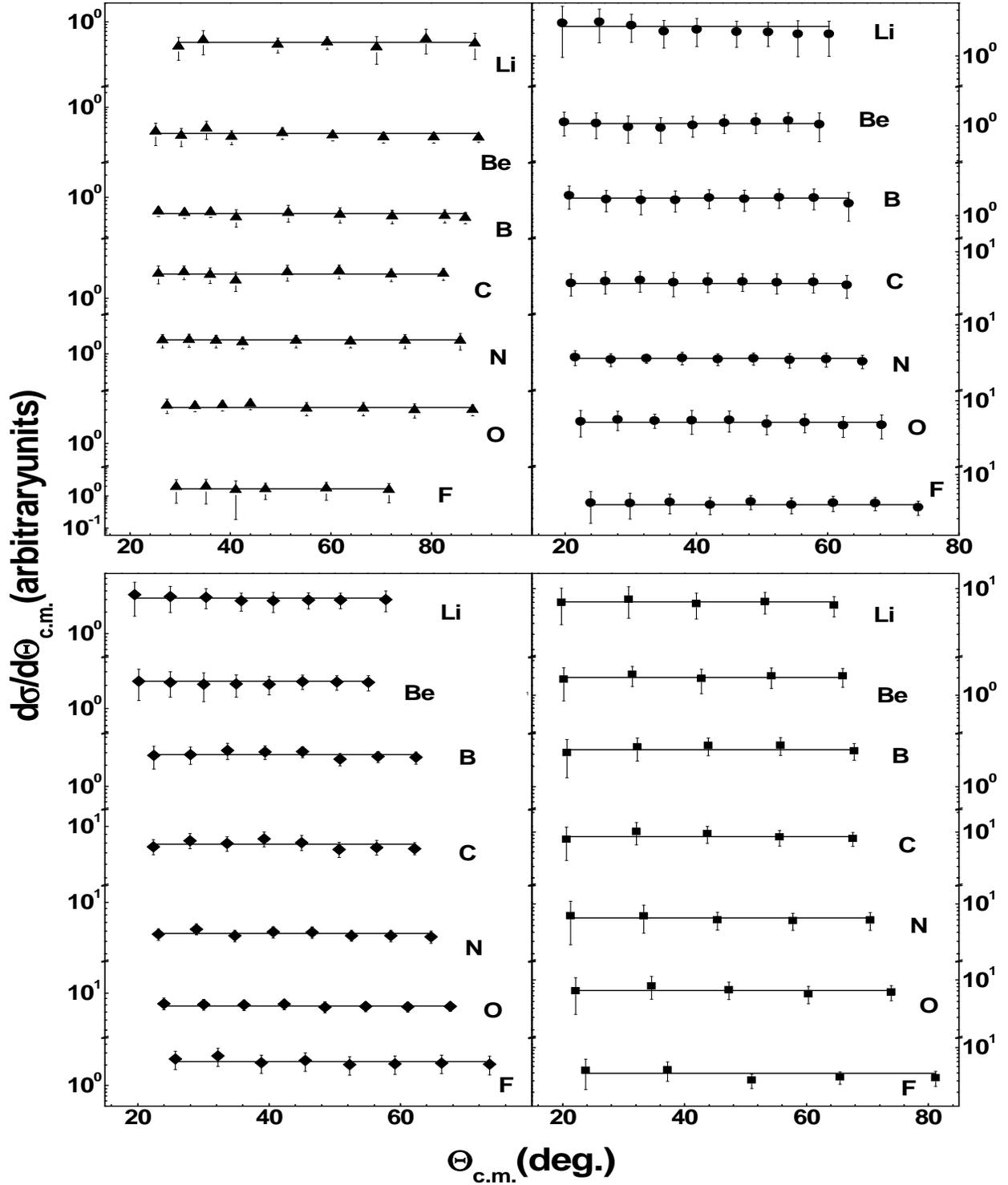,width=17.0cm,height=20.0cm}}
\caption{The c.m. angular distributions of the fusion-fission component for different fragments at the bombarding
energies 145 MeV (triangle), 158 MeV (circle), 200 MeV (diamond) and 218 MeV (square). The lines correspond
to fission-like angular distribution (d$\sigma$/d$\Omega$ $\sim$ 1/$\sin\Theta_{c.m.}$) fit to the data.}

\label{neal2}

\end{figure}

The  FF  and  the  DI components of the fragment angular distributions have
been obtained by  integrating  the  respective  energy  distributions.  The
center-of-mass (c.m.) angular distributions of the FF components of various
fragments  for all bombarding energies have been displayed as a function of
c.m. angle ($\Theta_{c.m.}$) in Fig.~\ref{neal2}. The  transformation  from
the  laboratory system to the c.m. system has been done with the assumption
of a two-body kinematics averaged over total kinetic energy  distributions.
The  angular  distributions  of  FF  components exhibit d$\sigma$/d$\Omega$
$\sim$   1/$\sin   \Theta_{c.m.}$-like   dependence   (solid    lines    in
Fig.~\ref{neal2}),  which are in conformity with the systematics of fission
decay of a fully equilibrated system.

\begin{figure}

{\epsfig{file=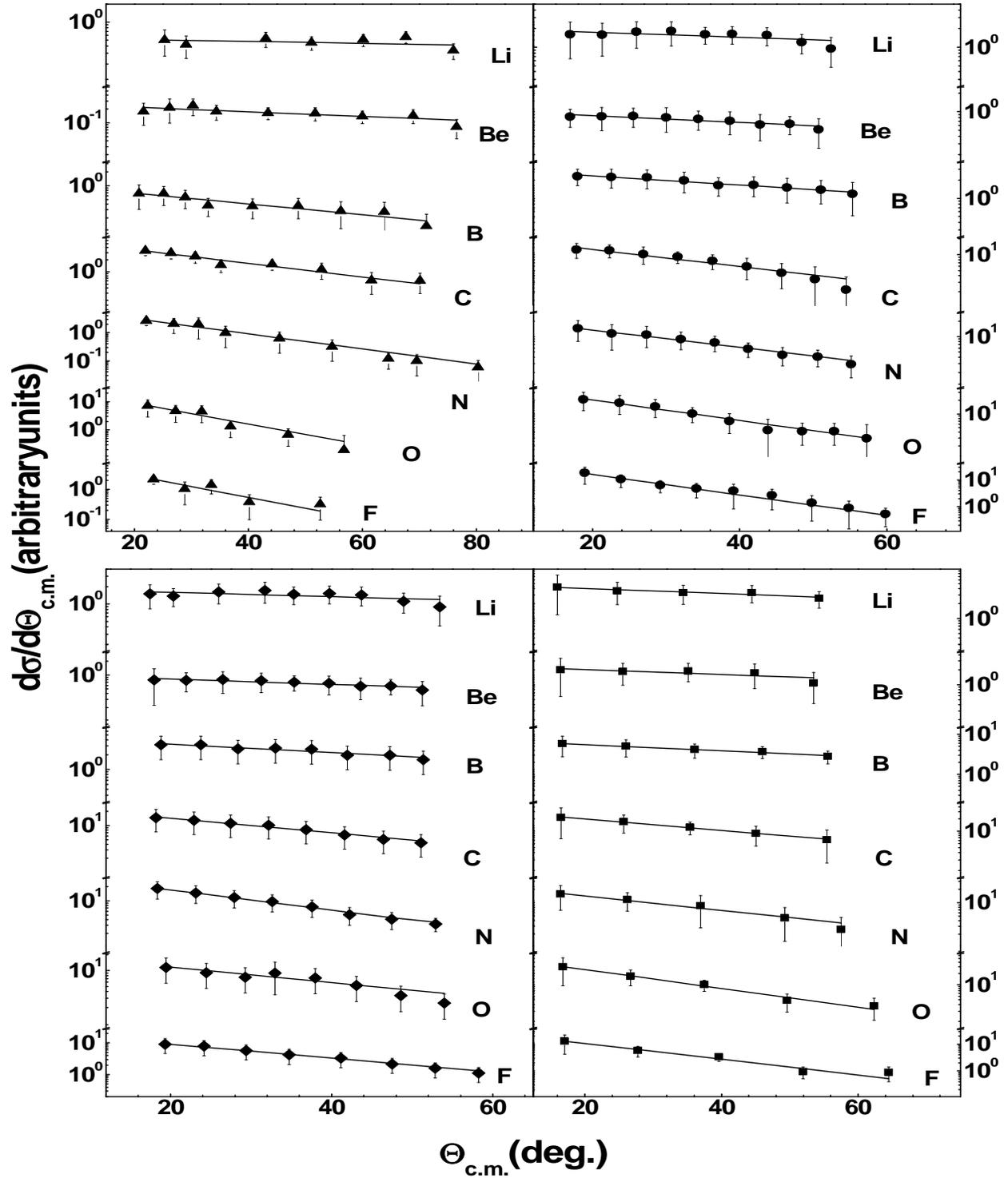,width=17.0cm,height=20.0cm}}
\caption{The c.m. angular distributions of the deep-inelastic component for different fragments at the bombarding
energies 145 MeV (triangle), 158 MeV (circle), 200 MeV (diamond) and 218 MeV (square). The lines are
exponential fit [Eq.~\ref{eq1}] to the data.}

\label{neal3}

\end{figure}

The  c.m. angular distributions of the DI components of different fragments
for all bombarding energies have been displayed as a function of c.m. angle
in Fig.~\ref{neal3}. A rapid fall of the angular distribution, (faster than
d$\sigma$/d$\Omega$ $\sim$ 1/$\sin  \Theta_{c.m.}$),  indicates  a  shorter
lifetime  of the composite system. Such lifetimes are incompatible with the
formation of an  equilibrated  compound  nucleus,  but  may  still  reflect
significant  energy  damping  within  a  deep-inelastic mechanism. From the
measured forward peaked angular distribution, it is  possible  to  estimate
the  lifetime  of  the  intermediate  dinuclear complex using a diffractive
Regge-pole model \cite{be4,mik80}. The angular distributions  have  been
fitted with the following expression:

\begin{equation}
\frac{d\sigma}{d\Omega}         \propto         \frac{C}{\sin\Theta_{c.m.}}
e^{-\Theta_{c.m.} / \Theta_{o}}, \label{eq1} \end{equation}

where,  $\Theta_{o}$  is  called  the  ``life angle", which is the angle of
rotation of the dinuclear composite during the time  interval  between  its
formation  to  the  decay  into  two  fragments.  The fit to the DI angular
distribution with Eq.~\ref{eq1} has been shown in  Fig.~\ref{neal3}  (solid
line).  The  values  of  $\Theta_{o}$  thus  obtained  are  given  in Table
\ref{tbl1}.

\subsection{Average Q-value distribution}

\begin{figure}

{\epsfig{file=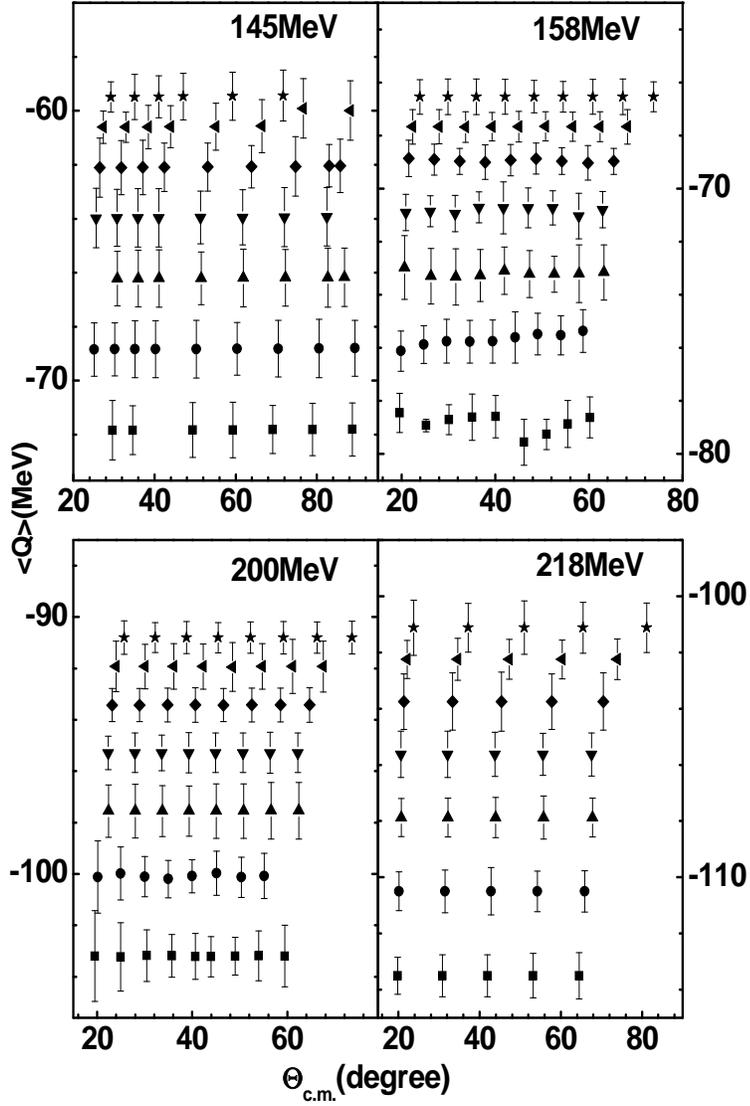,width=10.0cm,height=15.0cm}}
\caption{The average Q-values of the fusion-fission component for the fragments
Li(square), Be(circle), B(triangle), C(inverted triangle),
N(diamond), O(left triangle) and F(star) at each bombarding energy.}

\label{neal4}

\end{figure}

\begin{figure}

{\epsfig{file=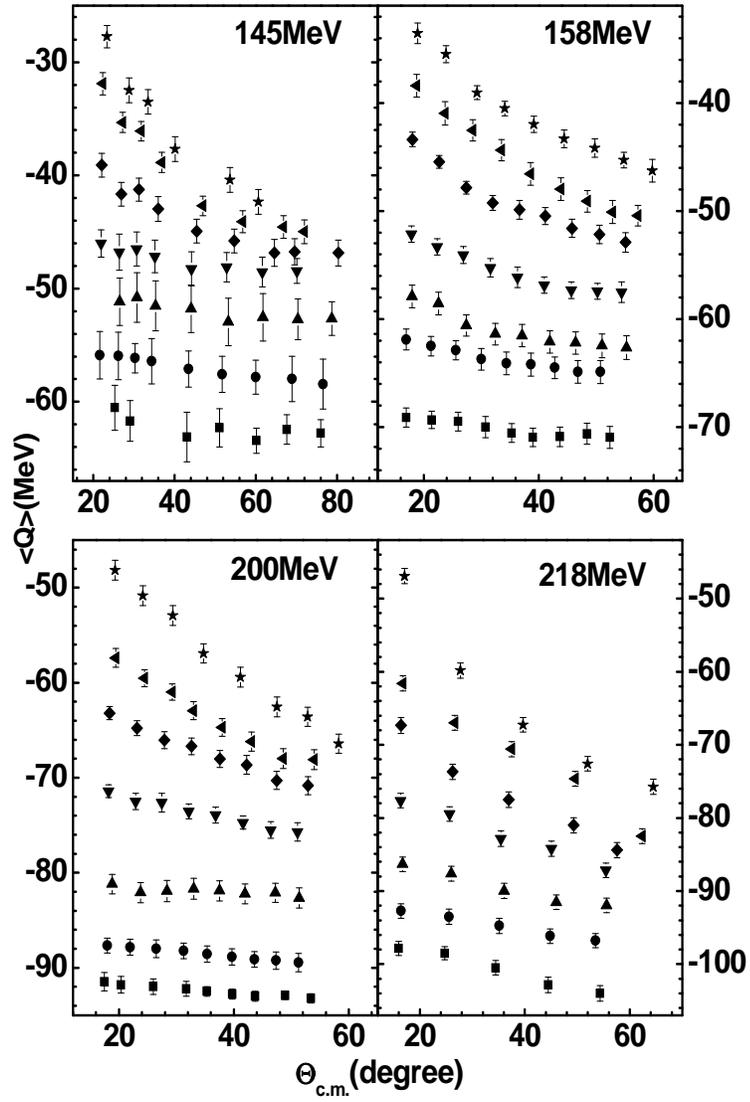,width=10.0cm,height=15.0cm}}
\caption{Same as Fig. \ref{neal4} for the deep-inelastic component.}

\label{neal5}

\end{figure}

The  average  Q-value,  $<Q>$,  has  been  estimated from the total average
kinetic  energy  of  the  fragments,  $E_{K}^{tot}$,  using  the  relation,
$E_{K}^{tot} = E_{c.m.} + <Q>$. The fragment total average kinetic energies
in  the  center-of-mass  have  been obtained from the respective laboratory
values  assuming  two  body  kinematics.  The  variations  of  $<Q>$   with
center-of-mass  emission  angle  for  the  fragments  (3 $\leq$ Z $\leq$ 9)
obtained  at  different  bombarding  energies  have   been   displayed   in
Figs.~\ref{neal4}~and~\ref{neal5}.   The  fragment  kinetic  energies  were
appropriately corrected for particle evaporation from the  excited  primary
fragments  assuming  thermal equilibrium of the dinuclear composite system.
It is observed that for all the fragments  at  all  energies,  the  average
Q-values   corresponding   to   FF   components   are  independent  of  the
center-of-mass emission angles, as  expected  [see  Fig.~\ref{neal4}].  The
average Q-values for DI components have been displayed in Fig.~\ref{neal5}.
The $<Q>$ values for DI components for the fragments Li, Be and B are found
to  be  nearly constant as a function of angle for all bombarding energies,
whereas those  for  C  to  F  are  found  to  decrease  at  forward  angles
($\Theta_{c.m.}$  $\lesssim$  40$^{o}$)  and  then  they  gradually tend to
become constant; these imply that, beyond this point,  the  kinetic  energy
damping is complete and dynamic equilibrium has been established before the
scission of the dinuclear composite takes place.

\subsection{Average velocity}

The  average  velocities  of  the  FF  component of the fragments have been
computed from their respective centroid energies. The mass number, $A$,  of
the  fragments  have  been  estimated  from  the  respective experimentally
obtained Z values using the empirical relation \cite{charity}:

\begin{equation}
A = Z \times (2.08 + 0.0029 \times Z).
\label{z}
\end{equation}

\begin{figure}

{\epsfig{file=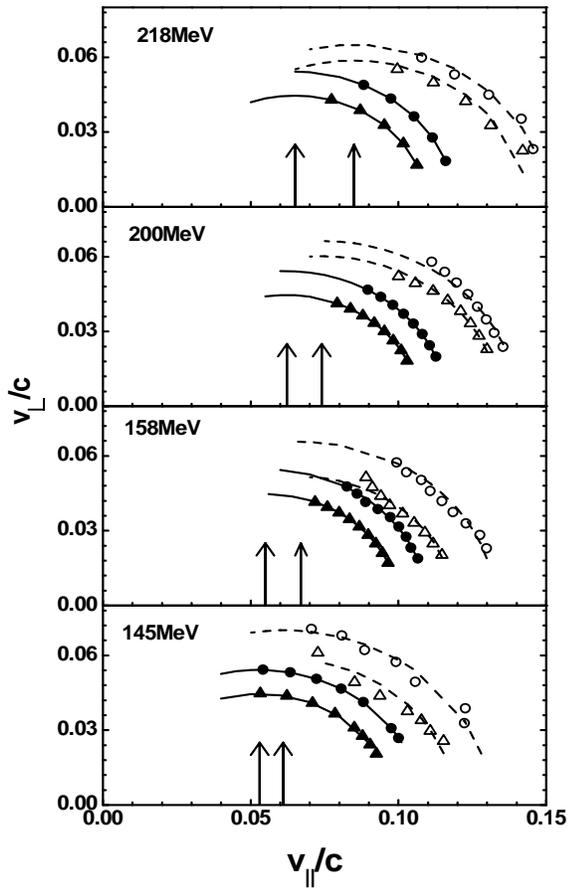,width=8.0cm,height=12.0cm}}
\caption{The average velocities of the fragments Lithium (circle) and Oxygen (triangle) plotted in
$v_\parallel - v_\perp$ plane at different bombarding energies for FF (solid symbol) and DI (open symbol)
components. The arrows correspond to the source velocities, $v_{CN}$ (lower) and $v_{DI}$ (higher).
The circles correspond to the most probable velocities for FF (solid) and DI (dashed) components.}

\label{neal6}

\end{figure}

The average velocities have been plotted in the $v_\parallel$ vs. $v_\perp$
plane  for  two  representative  fragments  (Li and O) in Fig.~\ref{neal6},
where the measured average velocities of the FF components were denoted  by
solid  symbols  and the corresponding $v_{CN}$ was represented by arrow. It
is seen that the average velocities fall on a circle (solid curve) centered
around $v_{CN}$, the compound nucleus velocity. This means that the average
velocities (as well as kinetic energies) of the fragments  are  independent
of  the c.m. emission angles at all incident energies considered here. This
clearly  indicates  that  these  fragments  are  emitted   from   a   fully
equilibrated  compound nucleus emission source with full momentum transfer.
The magnitude of the average fragment velocities ({\it i.e.}, the radii  of
the  circles  in  Fig.~\ref{neal6}) decreases with the increase of fragment
mass, which is indicative of the binary nature of the emission.

The  average  velocities  of  the  DI  components  at  different bombarding
energies have also been plotted in Fig.~\ref{neal6} (open symbols) for  the
same  fragments  Li  and  O.  It  is found that for this component also the
average velocities of all fragments fall on a  circle  (dashed  curve)  but
centered around a higher velocity source ($v_{DI}$). The values of $v_{CN}$
and  $v_{DI}$ for all bombarding energies are given in Table \ref{tbl2}. It
is interesting to note that at each bombarding  energy,  all  DI  fragments
(only Li and O are shown in figure) are emitted from the same source moving
with  velocity  $v_{DI}$.  Thus  the emission of DI fragments, may also be,
like FF fragments, visualized in terms of  emission  from  an  equilibrated
intermediate velocity source.

\subsection{Total elemental yield}

\begin{figure}

{\epsfig{file=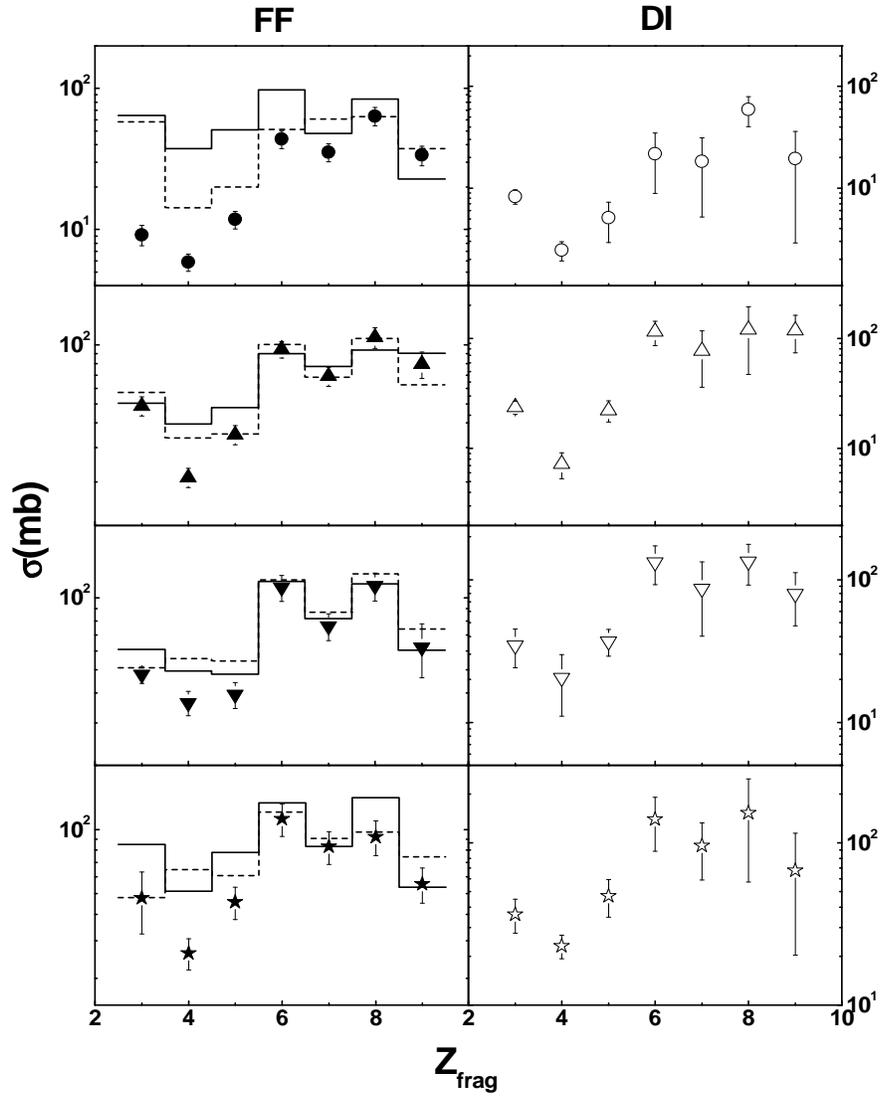,width=12.0cm,height=15.0cm}}
\caption{The total elemental cross-sections of FF(filled symbol) and DI(open symbol) components for
$^{20}$Ne + $^{27}$Al reaction at the bombarding energies 145 MeV (circle), 158 MeV (triangle),
200 MeV (inverted triangle) and 218 MeV (star), respectively, plotted as a function of fragment charge.
The histograms represent  the corresponding  theoretical predictions using CASCADE (solid) and EHFM (dashed).}

\label{neal7}

\end{figure}

The total fusion-fission ($\sigma_{FF}$) and deep-inelastic ($\sigma_{DI}$)
cross-sections  for  different  fragments have been obtained by integrating
the respective double differential cross-section ($d^{2}\sigma/d\Omega dE$)
over the whole energy and angular range. The cross-sections  thus  obtained
for   different  fragments  at  different  bombarding  energies  have  been
displayed in Fig.~\ref{neal7} as a function of  fragment  charge  Z.  Total
uncertainties  in the estimation of $\sigma_{FF}$ and $\sigma_{DI}$ are due
to the experimental threshold and the limited angular  range  of  the  data
(error  bars  in  Fig.~\ref{neal7}).  The  total elemental yields of the FF
components  ($\sigma_{FF}$)  have  been  compared  with   the   theoretical
estimates  of  the  same  obtained  from the statistical model code CASCADE
\cite{pul},  and,  from  the   extended   Hauser-Feshbach   method   (EHFM)
\cite{Matsuse97}.  The  model calculations have been performed by using the
critical angular momentum value for the respective bombarding  energy.  The
experimental fragment emission cross-sections for FF component are shown in
Fig.~\ref{neal7}  (left) by filled symbols and the theoretical estimates of
the same are represented by solid (CASCADE) and dashed  (EHFM)  histograms.
It  is  seen  from  the figure that the theoretical predictions are in fair
agreement with the experimental results except for Z =  4  fragment,  where
the  experimental  values  are much smaller than the respective theoretical
estimates; this might be because  of  non-detection  of  particle  unstable
$^{8}$Be,  which  decays  (into  two $\alpha$-particles) almost immediately
after production and thus escapes detection. The total elemental  yield  of
the  DI  component ($\sigma_{DI}$) are shown in Fig.~\ref{neal7} (right) by
open symbols. It has been found  that  a  large  fraction  of  the  heavier
fragment  (C,N,O,F) yield is due to the DI mechanism for all the bombarding
energies. The ratio of DI cross-section to FF cross-section increases  with
bombarding  energy (Table \ref{tbl3}). However, the FF and DI processes are
comparable in this bombarding energy range.

\section{Discussion and Conclusion}

\subsection{Time scale}

The  time scale of DI process can be estimated from DI angular distribution
using Eq.~\ref{eq1}, which describes the decay of a rotating dinucleus with
an  angular  velocity  $\omega$  =  $\hbar  l$/$\mu  R^{2}$,  where   $\mu$
represents  the  reduced  mass  of  the system, $l$ is the angular momentum
[$l_{cr} < l  \lesssim  l_{gr}$;  $l_{cr}$,  $l_{gr}$  being  the  critical
angular   momentum   for   fusion   and   the   grazing  angular  momentum,
respectively], $R$ represents the distance between the two centers  of  the
dinucleus,  and  $\tau$  is  the  time interval during which the two nuclei
remain in solid contact in the form of the rotating dinucleus.  The  ``life
angle"  ($\Theta_{o}$)  is  then the product of angular velocity ($\omega$)
and the rotation time ($\tau$). The characteristics of a  reaction  process
depends  on  the  value of $\Theta_{o}$. Smaller values of $\Theta_{o}$ are
associated with  faster  processes  for  which  the  corresponding  angular
distributions are more forward peaked. Large values of $\Theta_{o}$ ($\geq$
2$\pi$)   are  associated  with  slow  processes  with  lifetime  large  or
comparable to the dinucleus rotation period $\tau_{rot}(=2\pi/\omega)$, the
value    of    which     lies     typically     in     the     range     of
$\sim$~1-2~$\times$~10$^{-21}$    sec.    In    these   cases,   long-lived
configurations are assumed to be formed and the angular distributions  tend
to become symmetric around $90^{o}$ in the c.m. (d$\sigma$/d$\Omega$ $\sim$
1/$\sin  \Theta_{c.m.}$  type  distribution).  The  FF  process  is  thus a
limiting case of DI process,  where  a  very  long-lived  configuration  is
assumed to be formed and the angular distribution becomes $\propto$ 1/$\sin
\Theta_{c.m.}$.  The  c.m.  angular distributions of DI component have been
fitted with Eq.~\ref{eq1} and the time scales thus obtained  are  given  in
Table  \ref{tbl1}  for  different  fragments  emitted  in  the  $^{20}$Ne +
$^{27}$Al  reactions.  The upper (lower) limit of $\tau$ corresponds to the
estimate with $l$ = $l_{cr}$ ($l_{gr}$). The values of time
scales are found to vary in the range of 10$^{-21}$ -- 10$^{-22}$ sec.,
depending  on  the  bombarding  energy and fragment mass. It has been found
that the time scale decreases as the fragment charge increases.  For  lower
mass  fragments  (Li  to  B),  the time scale decreases with the bombarding
energy also and for other fragments (C to F) it is  almost  independent  of
bombarding  energy. This is expected because the heavier fragments (near to
the projectile) requires less number  of  nucleon  transfer  and  therefore
lesser  time. On the other hand, the emission of lighter fragments requires
more number of nucleon exchanges and therefore longer time.

\subsection{Angular momentum dissipation}

\begin{figure}

{\epsfig{file=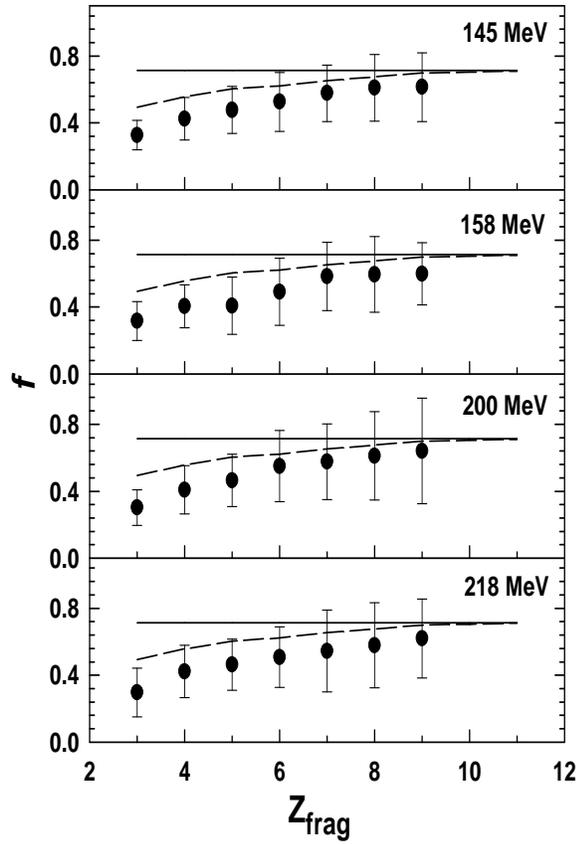,width=8.0cm,height=12.0cm}}
\caption{The angular momentum dissipation factors extracted for different fragments (filled circles) at various
incident energies. The corresponding empirical  limits are shown by solid (rolling) and dashed (sticking)  lines, respectively.}

\label{neal8}
\end{figure}

In  addition  to  kinetic  energy  dissipation,  the  dissipative heavy ion
collision processes also result  in  significant  dissipation  of  relative
angular  momentum  in the entrance channel. Phenomenologically, the kinetic
energy dissipation originates from the radial and tangential  component  of
friction  between  the  surfaces  of  the rotating dinuclear system; On the
other hand, the angular momentum  dissipation  is  decided  solely  by  the
tangential  component  of the friction, and the magnitude of dissipation is
expected to lie within any of the two phenomenological limits (rolling  and
sticking).  However, very large dissipation of relative angular momentum in
excess of the sticking limit predictions has  also  been  reported  in  the
literature  \cite{bh04,sh821}.  This  anomaly  may  be  due  to  the
ambiguity in  the  determination  of  the  magnitude  of  angular  momentum
dissipation  (and  vis-$\grave{a}$-vis  the  rotational contribution to the
fragment kinetic energy). Moreover, the estimation of the angular  momentum
in  the exit channel is strongly dependent on the scission configuration of
the rotating dinuclear system. An independent estimation  of  the  scission
configuration  is  necessary  to  estimate  the  angular  momentum transfer
properly. Generally, it is estimated from the total kinetic energy  of  the
rotating dinuclear system, $E_{K}^{tot}$, which is given by,

\begin{equation}
E_K^{tot} = V_N(d) + f^2 {\hbar^2 l_i (l_i + 1) \over 2 \mu d^2} ,
\label{dis}
\end{equation}

\noindent  where  $V_N(d)$  is  the  contribution  from Coulomb and nuclear
forces at dinuclear separation distance $d$, $\mu$ is the reduced  mass  of
the  dinuclear configuration, $l_i$ is the relative angular momentum in the
entrance channel and $f$ is the numerical factor denoting the  fraction  of
the  angular  momentum  transferred depending on the strength of tangential
friction. In absence of any method to estimate the values of  $f$  and  $d$
independently,  in  an  earlier  investigation  \cite{nato,van83}  $f$  was
assumed to be equal to its limiting value (corresponding  to  the  sticking
limit)  and  whole  of the fragment yield was assumed to be of DI origin to
arrive at an extended dinuclear configuration  for  $^{20}$Ne  +  $^{27}$Al
system.  However,  we  have  demonstrated  [Ref.~\cite{b2,bh04} and
present paper] that, at these energies, a significant part of the  fragment
yield is of FF origin and therefore this part should be subtracted from the
total  yield  to  properly  estimate  the DI yield. The extracted FF and DI
yields can be utilized to estimate the values of  $d$  and  $f$.  A  simple
procedure    for    estimating    both   $d$   and   $f$   was   given   in
Ref.~\cite{bh04}. Deep  inelastic  collisions  are  believed  to  occur
within  the  angular  momentum window between the critical angular momentum
for fusion, $l_{cr}$  and  the  grazing  angular  momentum,  $l_{gr}$.  The
partially  dissipative part of it (DI at forward angles) originates in near
peripheral collisions ($l \sim l_{gr}$), which correspond to small  overlap
and  a  fairly  elongated dinuclear configuration; On the other hand, fully
energy equilibrated dissipative components (at larger angles) correspond to
more compact collisions near $l \sim l_{cr}$. Moreover, the  fusion-fission
yield  is  also most predominant in the vicinity of $l \sim l_{cr}$. It is,
therefore,  likely  that  the  exit  channel  configurations  of  both  the
processes  are  similar and it appears to be reasonable to assume a compact
scission shape for the fully energy damped component of the  DI  yield.  In
the  present  work,  we  estimated  the  scission  configuration  from  the
extracted fusion-fission component of the measured fragment energy spectra.
The separation distance $d$ between the two fragments at the scission point
is calculated from the energy centroid of the FF energy spectra.  The  mean
values of $d$ thus estimated are 7.7 $\pm$ 1.2 fm for $^{20}$Ne + $^{27}$Al
reaction. Assuming these scission configurations, Eq.~\ref{dis} may then be
used  to  extract the angular momentum dissipation factor, $f$, in the case
of fully energy damped DI collisions.  The  values  of  $f$  extracted  for
different  energies  for  $^{20}$Ne  +  $^{27}$Al reaction are displayed in
Fig.~\ref{neal8} alongwith the rolling (solid line)  and  sticking  (dashed
line)  limit  predictions for the same. During the calculation the value of
initial angular momentum $l_i$ was  taken  to  be  equal  to  the  critical
angular momentum for fusion, $l_{cr}$.

\begin{figure}

{\epsfig{file=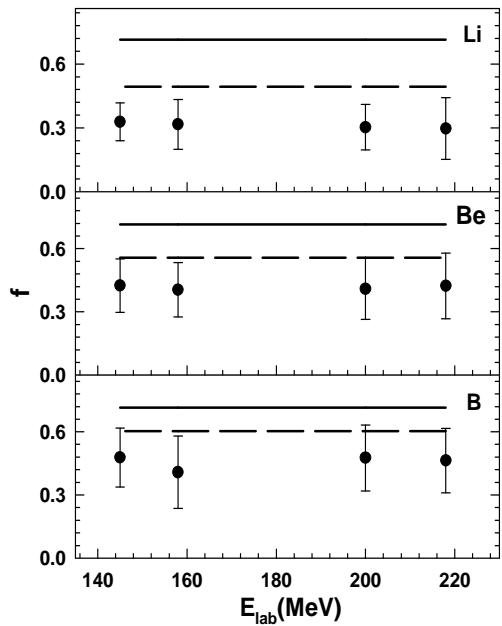,width=7.0cm,height=8.5cm}}
\caption{The variation of angular momentum dissipation factor as a function of incident energy. Filled circles are
the experimental estimates; empirical rolling and sticking limits for the same are represented by solid and dashed lines,
respectively.}

\label{neal9}
\end{figure}

It is apparent from Fig.~\ref{neal8} that for all the reactions considered,
there  is  discrepancy  between the experimental and empirical estimates of
angular momentum dissipation; so far as the
lighter fragments (Z = 3 -- 5) are concerned. For these fragments,
the  experimental  estimates  of angular momentum dissipation are more than
their limiting values predicted under the rolling and sticking  conditions.
The discrepancy is more for low mass fragments, and gradually decreases for
heavier  fragments.  This may be qualitatively understood as follows: it is
known from the study of dissipative dynamics of fission \cite{dhara}  that,
strong   frictional   forces   in  the  exit  channel  causes  considerable
retardation of the scission process leading to increase  in  scission  time
scale.  As  the  the  exit  channel  configurations  of the fully damped DI
process are taken to be similar to those for FF process  (except  that  the
dinuclear system, in case of DI collision, is formed beyond the conditional
saddle  point  directly),  the  dynamics  of DI process may also experience
stronger frictional forces. Microscopically, friction is generated  due  to
stochastic  exchange  of nucleons between the reacting partners through the
window formed by the overlap of  the  density  distributions  of  the  two.
Stronger  friction,  in  this  scenario, essentially means larger degree of
density  overlap  and  more  nucleon  exchange.  Consequently,  lighter  DI
fragments  (corresponding  to  more  net  nucleon  transfer) originate from
deeper collisions, for which interaction times are larger.  Therefore,  the
angular   momentum  dissipation,  originating  due  to  stochastic  nucleon
exchange, may also be more which,  at  least  qualitatively,  explains  the
observed  trend. The angular momentum dissipation for a particular fragment
({\it e.g.}, Li, Be, B) is found to be nearly independent of the bombarding
energy. This is clearly shown from Fig.~\ref{neal9} where  the  factor  $f$
have  been displayed as a function of bombarding energy for the $^{20}$Ne +
$^{27}$Al reaction. This  may  be  further  indicative  of  the  stochastic
nucleon exchange origin of the frictional force for the fully energy damped
DI  process,  for which scission configuration is nearly independent of the
bombarding energy.

To  summarise,  We  have  studied  $^{20}$Ne  (145,  158,  200,  218 MeV) +
$^{27}$Al reactions and extracted the contributions to the  fragment  yield
from   fusion-fission  and  deep  inelastic  processes.  The  c.m.  angular
distributions  of  FF  component  was  found   to   have   $\sim$   1/$\sin
\Theta_{c.m.}$   dependence   whereas  those  of  DI  component  showed  an
exponential fall off at forward angles. The time scale of  the  DI  process
has  been  estimated  form  the DI angular distribution. The lifetime of DI
process has been found to decrease with increasing fragment mass  and  also
with  increasing  bombarding  energy.  The  fusion-fission component of the
emitted fragments has been found to be originate from the compound  nucleus
source  (moving with velocity $v_{CN}$), while the deep inelastic component
of the fragments are found to be  emitted  from  an  intermediate  velocity
source having velocity $v_{DI}$, which is higher than $v_{CN}$. The average
Q-values  for  DI  component  have  been found to decrease with increase of
emission angles and saturate  at  higher  angles  signifying  fully  energy
damped  process  at these angles; On the other hand, those for FF component
have been found to  be  independent  of  emission  angles  as  expected  in
equilibrium  emission.  The  elemental cross-sections have been obtained by
integrating separately the energy distributions of the FF and DI components
over the corresponding energies and  over  the  whole  angular  range.  The
fusion-fission  fragment yield, $\sigma_{FF}$, have been found to be fairly
well explained in terms of statistical model. The $\sigma_{DI}/\sigma_{FF}$
value increases with bombarding energies,  which  is  expected  because  of
increasing  contribution  of  DI  processes  at higher energies. Assuming a
compact exit channel configuration (estimated from extracted FF part of the
spectra) for the fully damped part of  the  deep-inelastic  reactions,  the
angular  momentum  dissipation  has been estimated and it has been found be
more than the corresponding phenomenological  limits.  The  deviations  are
found  to  be  more  for  lighter  fragments,  which  may be related to the
microscopic (stochastic nucleon exchange) origin of nuclear friction.

\acknowledgements

The  authors  like to thank the cyclotron operating crew for smooth running
of the machine, and H.~P.~Sil for the fabrication of thin silicon detectors
for the experiment. One of the authors (A.~D.) acknowledges with thanks the
financial support provided by the  Council  of  Scientific  and  Industrial
Research, Government of India.

\begin{table}
\caption{The  time scales for emission of different DI fragments. The upper
(lower) limit corresponds to $l_{cr}$ ($l_{gr}$). The numbers  in  brackets
denote corresponding uncertainties.}

\begin{tabular}{cccccc}
$E_{lab}$&$l_{cr}$&$l_{gr}$&Fragment&$\Theta_{o}$&$\tau$ \\
(MeV)&($\hbar$)&($\hbar$)&&(radian)&(10$^{-22}$ $s$) \\ \tableline
145&37&51&Li&7.68(1)&20.85(3) - 15.13(2) \\
&&&Be&1.99(1)&5.94(2) - 4.31(2) \\
&&&B&0.67(2)&2.16(7) - 1.57(5) \\
&&&C&0.43(1)&1.44(3) - 1.04(3) \\
&&&N&0.31(1)&1.10(3) - 0.80(2) \\
&&&O&0.22(1)&0.82(3) - 0.59(3) \\
&&&F&0.21(1)&0.82(4) - 0.60(2) \\ \tableline
158&38&54&Li&3.36(1)&8.88(3) - 6.25(2) \\
&&&Be&1.76(1)&5.11(3) - 3.60(2) \\
&&&B&1.10(1)&3.46(3) - 2.43(2) \\
&&&C&0.54(2)&1.76(6) - 1.24(4) \\
&&&N&0.47(1)&1.62(3) - 1.14(2) \\
&&&O&0.35(2)&1.26(8) - 0.89(5) \\
&&&F&0.18(2)&0.68(8) - 0.48(5) \\ \tableline
200&41&63&Li&3.09(1)&7.57(2) - 4.93(1) \\
&&&Be&1.59(5)&4.29(12) - 2.79(8) \\
&&&B&0.98(2)&2.85(6) - 1.85(4) \\
&&&C&0.53(2)&1.60(6) - 1.04(4) \\
&&&N&0.36(1)&1.15(3) - 0.75(2) \\
&&&O&0.32(2)&1.07(6) - 0.70(4) \\
&&&F&0.24(3)&0.84(9) - 0.55(7) \\ \tableline
218&42&66&Li&1.37(2)&3.28(4) - 2.09(3) \\
&&&Be&1.14(1)&3.00(2) - 1.90(2) \\
&&&B&0.99(1)&2.81(3) - 1.79(2) \\
&&&C&0.56(2)&1.65(6) - 1.05(4) \\
&&&N&0.52(1)&1.62(3) - 1.03(2) \\
&&&O&0.30(2)&0.98(6) - 0.62(3) \\
&&&F&0.24(1)&0.82(4) - 0.52(2) \\
\end{tabular}
\label{tbl1}
\end{table}

\begin{table}
\caption{The values of different velocity sources.}
\begin{tabular}{ccc}
$E_{lab}$&$v_{CN}$&$v_{DI}$ \\
(MeV)&(v/c)&(v/c) \\ \tableline
145&0.053$\pm$0.001&0.061$\pm$0.003 \\
158&0.055$\pm$0.001&0.067$\pm$0.002 \\
200&0.062$\pm$0.001&0.074$\pm$0.002 \\
218&0.065$\pm$0.001&0.085$\pm$0.003 \\
\end{tabular}
\label{tbl2}
\end{table}

\begin{table}
\caption{The $\sigma_{DI}$/$\sigma_{FF}$ values for different fragments at all bombarding energies.}
\begin{tabular}{ccccc}
$E_{lab}$&\multicolumn{4}{c}{$\sigma_{DI}$/$\sigma_{FF}$}\\
(MeV)&C&N&O&F\\ \tableline
145&0.51$\pm$0.10&0.52$\pm$0.11&0.93$\pm$0.17&0.58$\pm$0.12 \\
158&1.20$\pm$0.24&1.11$\pm$0.22&1.11$\pm$0.22&1.47$\pm$0.29 \\
200&1.21$\pm$0.23&1.14$\pm$0.23&1.20$\pm$0.24&1.28$\pm$0.26 \\
218&1.25$\pm$0.25&1.15$\pm$0.23&1.65$\pm$0.33&1.22$\pm$0.23 \\
\end{tabular}
\label{tbl3}
\end{table}

\newpage

\vspace{0.8cm}

\vspace{0.4cm}

\end{document}